\documentclass[preprint2]{aastex}% preprint2 produces a double-column, single-spaced document:
\usepackage{natbib}
\shorttitle{Radial Stellar Pulsation and Three-dimensional Convection. III.}
\shortauthors{Geroux and Deupree}

\begin{document}

\title{RADIAL STELLAR PULSATION AND THREE-DIMENSIONAL CONVECTION. IV. FULL AMPLITUDE THREE-DIMENSIONAL SOLUTIONS}

\author{Christopher M. Geroux\altaffilmark{1} and Robert G. Deupree}
\affil{Institute for Computational Astrophysics and Department of Astronomy and Physics, Saint Mary's University, Halifax, NS B3H 3C3 Canada}
\email{geroux@astro.ex.ac.uk}
\altaffiltext{1}{Now at Physics and Astronomy, University of Exeter, Stocker Road, Exeter, UK EX4 4QL}

\begin{abstract}
Three dimensional hydrodynamic simulations of full amplitude RR Lyrae stars have been computed for several models across the instability strip. The three dimensional nature of the calculations allows convection to be treated without reference to a phenomenological approach such as the local mixing length theory. Specifically, the time dependent interaction of the large scale eddies and the radial pulsation is controlled by the conservation laws, while the effects of smaller convective eddies are simulated by an eddy viscosity model. The light amplitudes for these calculations are quite similar to those of our previous two dimensional calculations in the middle of the instability strip, but somewhat lower near the red edge, the fundamental blue edge, and for the one first overtone model we computed. The time dependent interaction between the radial pulsation and the convective energy transport is essentially the same in three dimensions as it is in two dimensions. There are some differences between the light curves between the two and three dimensional simulations, particularly during decreasing light. Reasons for the differences, both numerical and physical are explored.
\end{abstract}


\section{INTRODUCTION}
Convection continues to be a major issue in the study of variable stars half a century after is was first suspected as being significant near the red edge of the instability strip \citep{Christy-1966a,Cox-1966b}. Simple prescriptions for the time dependent interaction between convection and pulsation such as freezing in the convective flux during the pulsation cycle \citep{Tuggle-1973} and having the convective flux instantaneously adjust to the current pulsational structure \citep{Cox-1966b} failed to provide realistic behavior near the red edge. Time dependent mixing length approaches \citep[e.g][]{Stellingwerf-1982a,Stellingwerf-1982b, Stellingwerf-1984a,Stellingwerf-1984b,Stellingwerf-1984c,Kuhfuss-1986,Xiong-1989} have been used \citep{Gehmeyr-1992a,Gehmeyr-1992b,Gehmeyr-1993,Bono-1994,Bono-1997a,Bono-1997b,Marconi-2003,Marconi-2007} to produce a red edge, but the light curves near the red edge do not agree well with observations \citep{Marconi-2007} and it is generally concluded that the treatment of convection needs improvement \citep{Buchler-2009, Marconi-2009}.

An alternative approach is to allow the time dependent interaction of convection and pulsation to be determined by finite difference approximations to the conservation laws. Such calculations \citep{Deupree-1977a,Deupree-1977b} follow the largest scale convective eddies and treat the unresolved small scale eddies as a viscosity acting on the large scale flow.

\citeauthor{Deupree-1977a} allowed the computational mesh to move radially with the horizontal average of the radial velocities on each spherical surface. In principle, one can allow the mesh to move however one chooses; in practice, this specific algorithm allowed the very narrow hydrogen ionization region to be poorly simulated during parts of the pulsation cycle. It took approximately twenty periods for this to become significant, so \citeauthor{Deupree-1977a} was able to compute pulsational growth rates, but not full amplitude solutions.  Based on the fact that 1D Lagrangian calculations allow the computation of full amplitude solutions, \citet[hereinafter GD1]{Geroux-2011} developed a 1D, 2D, and 3D radiation hydrodynamics code which makes the radial coordinate move so that the total amount of mass in a given spherical shell does not change during the calculation. Of course, this motion is Lagrangian in a 1D code, but not in multiple dimensions because material can flow in and out of a spherical shell; there can just be no net flow.

This approach does allow the calculation of full amplitude solutions. \citet[hereinafter GD2]{Geroux-2013b} computed full amplitude solutions in 2D for several models across the instability strip. They found the same time dependent relationship between convection and pulsation as found by Deupree (1977b). The pulsation amplitudes strongly decreased as the models approached the red edge, but models beyond the red edge encountered difficulties because the convection zone wanted to penetrate well below the hydrogen and helium ionization regions. More 2D calculations have become available \citep[e.g.,][]{Gastine-2011,Gastine-2008a,Gastine-2008b,Mundprecht-2013}, but the computational difficulties of the pulsation--convection interaction have led to a difference in emphasis between these and the current work. Those 2D simulations tend to focus on highly zoned calculations to obtain great detail about the convective behaviour. This necessarily limits the time scale over which the calculations can be performed and requires some other restrictive assumptions. Conversely our work accepts relatively modest zoning (particular angular) in the interests of being able to integrate over the many time steps necessary to obtain the full amplitude models which can be compared directly with observations.

Of course the turbulent nature of convective flow encountered near the surface in RR Lyrae variables is inherently 3D. \citet{Deupree-1977a} and GD2 argue that the important feature is the time dependent interaction between convection and pulsation, and not the details of the convective flow. This is, of course, debatable, but \citet[herinafter GD3]{Geroux-2014} have shown that this appears to be true, at least in a limited set of circumstances. 

In this paper we have carried a number of 3D calculations to full amplitude with the primary objective of comparing the results with the 2D full amplitude solutions in GD2. The physics in both sets of calculations are the same: the OPAL opacities \citep{Iglesias-1996} in conjunction with the low temperature \citet{Alexander-1994} opacities and the OPAL equation of state \citep{Rogers-1996}.  Radiation is treated with the diffusion approximation. An eddy viscosity is used to simulate the effect of unresolved convective eddies on the larger scale flow.  There are approximately 150 radial zones in the models, while the number of angular zones is 20 in 2D and 20 x 20 in 3D. The total width in each horizontal dimension was 6$^\circ$.  The conservation laws are explicitly computed with the exception of the energy equation, which is solved implicitly with a Newton-Raphson technique. More specific details are provided in GD1, GD2, and GD3.

Sixteen processors were used in parallel for all the 3D calculations presented here. Our 2D calculations take a few weeks to get to full amplitude, while the 3D calculations required several months. The reason for this amount of time is that the models require so many time steps to reach full amplitude. Even with current computational capabilities, only a limited number of 3D models can be done feasibly, including having more limited zoning than one might desire. Thus, we have computed some more 2D models to assess the sensitivity of the results to various features.

We begin with a comparison of the pulsation amplitude as a function of effective temperature followed with a comparison of individual light curves.

\section{COMPARISON OF 2D AND 3D FULL AMPLITUDE MODELS}

\begin{deluxetable}{cccccccccc}
%\tabletypesize{\tiny}
%\tabletypesize{\scriptsize}
%\tabletypesize{\normalsize}
\tablecaption{2D and 3D Full Amplitude Models}
\tablenum{1}
\tablehead{
\colhead{$T_{\rm eff}$} &
\colhead{D} &
\colhead{$A_V$} &
\colhead{$L_{\rm conv.}/L_{\rm tot.}$} &
\colhead{$\phi_L$} &
\colhead{$\Delta \langle T\rangle / \langle T\rangle$} &
\colhead{$\phi_T$} &
\colhead{$v_{\rm conv.}$} &
\colhead{$\phi_v$} &
\colhead{$v_{\rm amp.}$} \\
\colhead{K} &
\colhead{} &
\colhead{mag} &
\colhead{} &
\colhead{} &
\colhead{} &
\colhead{} &
\colhead{km s$^{-1}$} &
\colhead{} &
\colhead{km s$^{-1}$}
}
\startdata
6300&2&0.56&$0.60\pm 0.06$&$0.70$&$0.63\pm 0.01$&$0.72$&$20\pm 2$&$0.69$&$65 \pm 8$\\
    &3&0.44&$0.58\pm 0.03$&$0.66$&$0.69\pm 0.01$&$0.69$&$28\pm 2$&$0.63$&$73 \pm 2$\\
    \\
6400&2&0.64&$0.65\pm 0.01$&$0.73$&$0.65\pm 0.02$&$0.76$&$20\pm 2$&$0.71$&$82 \pm 1$\\
    &3&0.66&$0.45\pm 0.02$&$0.66$&$0.71\pm 0.01$&$0.69$&$31\pm 1$&$0.64$&$85 \pm 1$\\
    \\
6500&2&0.75&$0.61\pm 0.04$&$0.67$&$0.66\pm 0.01$&$0.70$&$21\pm 1$&$0.66$&$92 \pm 1$\\
    &3&0.76&$0.34\pm 0.01$&$0.71$&$0.72\pm 0.01$&$0.74$&$31\pm 1$&$0.71$&$94 \pm 1$\\
    \\
6600&2&0.83&$0.52\pm 0.03$&$0.67$&$0.67\pm 0.01$&$0.71$&$22\pm 1$&$0.67$&$96 \pm 1$\\
    &3&0.84&$0.23\pm 0.02$&$0.66$&$0.69\pm 0.01$&$0.69$&$28\pm 1$&$0.65$&$99 \pm 1$\\
    \\
6700&2&1.01&$0.34\pm 0.04$&$0.65$&$0.62\pm 0.01$&$0.76$&$15\pm 1$&$0.73$&$106\pm 1$\\
    &3&0.86&$0.11\pm 0.01$&$0.63$&$0.63\pm 0.01$&$0.73$&$20\pm 3$&$0.71$&$96 \pm 1$\\
    \\
6900&2&0.46&$0.05\pm 0.01$&$0.76$&$0.78\pm 0.04$&$0.61$&$11\pm 2$&$0.77$&$56 \pm 1$\\
    &3&0.38&$0.05\pm 0.01$&$0.73$&$0.82\pm 0.08$&$0.69$&$11\pm 2$&$0.72$&$53 \pm 1$\\
\enddata
\label{t1}
\end{deluxetable}

\begin{figure}
\center
\plotone{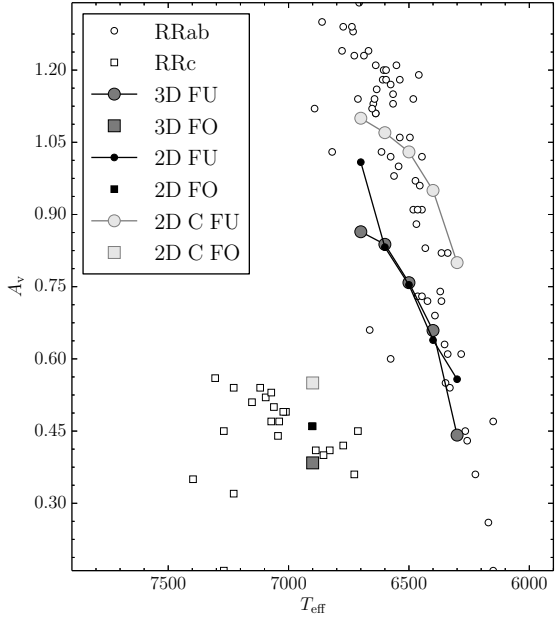}
\caption{Pulsation amplitude as a function of effective temperature in 2D, 3D, and 2D with double the resolution and double the angular extent. Squares indicate pulsation in the first overtone while circles indicate pulsation in the fundamental mode and open symbols indicate observations. The small black symbols show the low resolution and low extent 2D calculations (baseline case with 20 angular zones). The larger dark grey symbols show the 3D calculations and the large light grey symbols show the double extent and double resolution 2D calculations (Case C, discussed in section~4). The baseline 2D and 3D full amplitudes agree well in the middle of the instability strip, but less so elsewhere.}
\label{f1}
\end{figure}

We compare a few properties of the 2D and 3D models as functions of effective temperature in Table~\ref{t1}. These include the pulsation amplitude in the visual, the pulsation velocity amplitude, the maximum of the ratio of the convective luminosity to the total luminosity during the pulsation cycle, the maximum of the horizontal temperature variation compared to the horizontal average temperature, and the maximum convective velocity. Also included are the phases at which these maxima are found with phase zero being defined as the maximum in peak kinetic energy near maximum light. The errors in the phases are about 0.01, with all errors being determined over four consecutive periods. We see that the maximum convective luminosities are generally noticeably larger in 2D than in 3D, although we note the difficulty in comparing these because of the different geometrical balance between upward and downward flow in 2D and 3D (see CD3). This difficulty is further highlighted by the fact that the maximum horizontal temperature variation and convective velocity are generally higher in the 3D models nearer to the red edge. These maxima generally occur a little later (the 6500~K model is an exception) in 2D than in 3D, although the differences are sufficiently small as not to invalidate our contention that the time dependence is reasonably similar in 2D and 3D.

These similarities and differences appear in the light curves as well. We present the pulsation amplitude as a function of the effective temperature for our 2D and 3D models, along with the observed values from \citet{Cacciari-2005} for M3 in Figure~\ref{f1}. We first note that the pulsation amplitudes in the middle of the fundamental mode part of the instability strip ($T_{\rm eff}$=6400~K -- 6600~K) agree quite well for the 2D and 3D calculations.  For the models near both edges of the fundamental mode instability strip and for the first overtone model calculated, the 3D pulsation amplitudes are noticeably lower than those of the 2D models.  Before discussing these differences in amplitude in detail, we turn to a comparison of the individual light curves. 

\begin{figure}
\center
\plotone{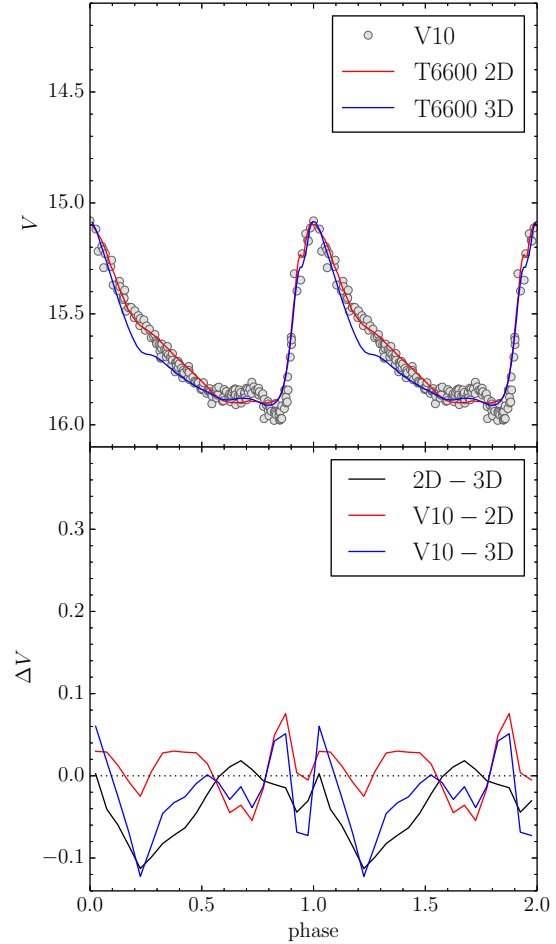}
\caption{Comparison of light curves from 2D and 3D simulations for an effective temperature of 6600~K and the light curve of V10 in M3 (top). The differences between the light curves at each phase are also shown (bottom).}
\label{f2}
\end{figure}

\begin{figure}
\center
\plotone{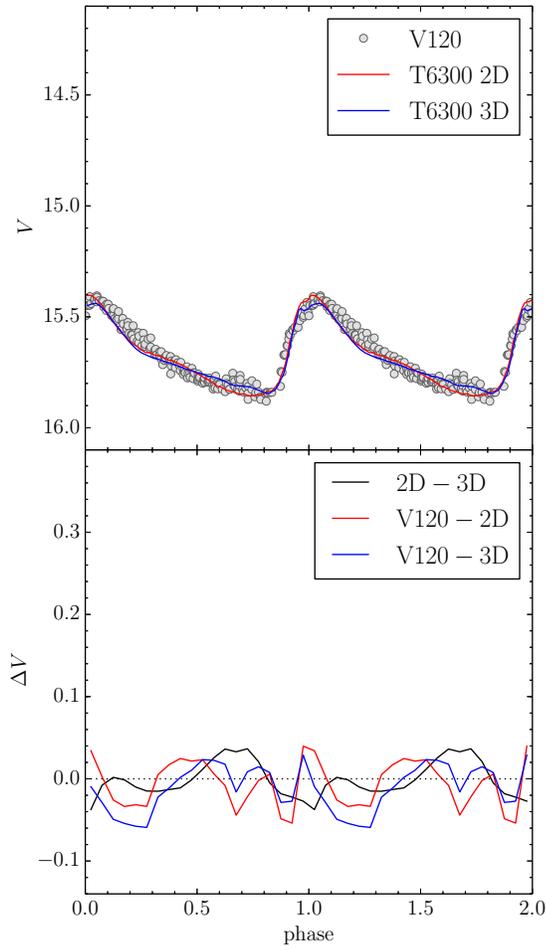}
\caption{Same as figure~\ref{f2}, except for an effective temperature of 6300~K. The observed variable is V120 in M3.}
\label{f3}
\end{figure}

\begin{figure}
\center
\plotone{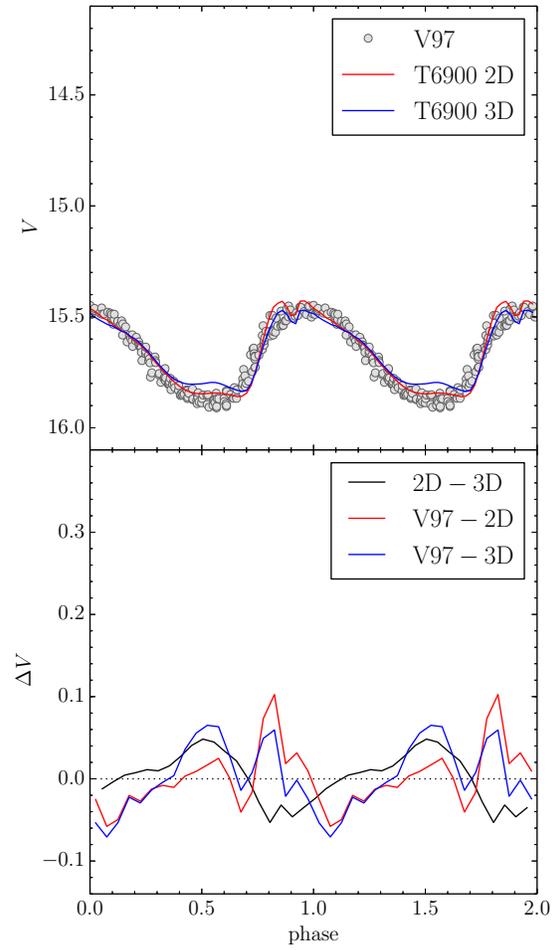}
\caption{Same as figure~\ref{f2}, except for the first overtone mode with an effective temperature of 6900~K. The observed variable is V97 in M3.}
\label{f4}
\end{figure}

We have previously shown a comparison between the 2D and 3D light curves for $T_{\rm eff} = 6500$~K (GD3). The differences in the computed light curves are very similar to the result for $T_{\rm eff}=6600$~K, shown in Figure~\ref{f2}: the 2D and 3D curves are essentially the same during rising light, but the 3D curve falls off more rapidly from peak light and then falls more slowly as minimum light is approached. In the 6500K case we noted that the 2D light curve resembles the observed light curves more closely during decreasing light. This is also true for the 6600~K case as well, as the comparison with the light curve of V10 in M3 \citep{Cacciari-2005} shows. Comparisons of the 2D light curves with observations are presented in GD2.

The same trend is also true for the $T_{\rm eff} = 6300$~K model, although the difference in amplitude between the 2D and 3D calculations somewhat masks the effect. We show this in Figure~\ref{f3}, along with the light curve of V120. Again, the 2D light curve provides marginally better agreement with the observations than the 3D light curve. One should keep in mind that some details of the light curves change as the amplitude changes.

A somewhat different picture is presented in Figure~\ref{f4} for the $T_{\rm eff}=6900$~K case. Here the 3D light curve falls more slowly from maximum light than the 2D light curve, and the two curves have nearly the same slope in the latter half of decreasing light until minimum light is approached. There is little discernible difference during rising light except near maximum light, which may be attributed to the lower amplitude of the 3D calculation. The comparison with the low amplitude first overtone pulsation V97 shows that the 3D calculation has a steeper rise to maximum light and remains at maximum light for a shorter time. The 2D light curve comparison with V125 shows the same trend (Figure 13 of GD2), but perhaps in a less pronounced way. We should also note that the difference in total amplitude may play a role in the apparent differences of the light curve shape as some features of the light curves shape can change with amplitude.

Thus, there are some general differences between the light curves of the 2D and 3D calculations which appear to occur over most of the instability strip. We now examine possible sources, both numerical and physical, for the differences.

\section{POSSIBLE ORIGINS OF DIFFERENCES BETWEEN 2D AND 3D CALCULATIONS}
\label{sec:2D-v-3D-diff}

\begin{deluxetable}{cccccccccc}
%\tabletypesize{\normalsize}
\tablecaption{Eddy Viscosity Parameter Study of 6500~K Model}
\tablenum{2}
\tablehead{
\colhead{EV} &
\colhead{D} &
\colhead{$A_V$} &
\colhead{$L_{\rm conv.}/L_{\rm tot.}$} &
\colhead{$\phi_L$} &
\colhead{$\Delta \langle T\rangle / \langle T\rangle$} &
\colhead{$\phi_T$} &
\colhead{$v_{\rm conv.}$} &
\colhead{$\phi_v$} &
\colhead{$v_{\rm amp.}$} \\
\colhead{} &
\colhead{} &
\colhead{mag} &
\colhead{} &
\colhead{} &
\colhead{} &
\colhead{} &
\colhead{km s$^{-1}$} &
\colhead{} &
\colhead{km s$^{-1}$}
}
\startdata
0.17&2&0.75&$0.61\pm 0.04$&$0.67$&$0.66\pm 0.01$&$0.70$&$21\pm 1$&$0.66$&$ 92 \pm 1$\\
0.25&2&0.76&$0.37\pm 0.04$&$0.68$&$0.63\pm 0.01$&$0.68$&$17\pm 1$&$0.68$&$100 \pm 1$\\
0.11&3&0.78&$0.50\pm 0.01$&$0.72$&$0.73\pm 0.01$&$0.74$&$34\pm 2$&$0.71$&$ 86 \pm 1$\\
0.17&3&0.76&$0.34\pm 0.01$&$0.72$&$0.70\pm 0.01$&$0.74$&$31\pm 1$&$0.71$&$ 94 \pm 1$\\
0.25&3&0.72&$0.14\pm 0.01$&$0.57$&$0.66\pm 0.03$&$0.72$&$22\pm 1$&$0.68$&$103 \pm 1$\\
\enddata
\label{t2}
\end{deluxetable}

Given that the 3D pulsation growth rates were lower than the 2D growth rates (GD3), however marginally, perhaps it is not surprising that the pulsation amplitude is less in 3D than 2D for cases near the instability strip boundaries for the given modes. The pulsation amplitude for the $T_{\rm eff}=6700$~K 3D model is somewhat less than that for the 2D model. Part of this is due to the fact that the 3D model has a very low growth rate and may not have reached full amplitude even after extensive time integration, although we doubt that it would reach the 2D amplitude based on this current rate of growth. 

The 2D models near the red edge have a tendency to form deeper convection zones. Our experiments with the 2D angular zoning in section 4 indicate that this tendency depends on the angular zoning, being less likely as the angular resolution or angular extent is increased. Forming a deep convection zone means that reaching the full amplitude solution may take some time because of the changing thermal energy content of the outer parts of the model. This may not be reflected in the amplitude of the light curve or the magnitude of the peak kinetic energy per period except over many pulsation cycles. The full amplitude should thus be considered more uncertain and sensitive near the red edge.

Looking at the first overtone model, we note that the 2D first overtone calculation was actually begun in the fundamental and made the transition to the first overtone during the hydrodynamic simulation while the 3D calculation was begun in the first overtone. Furthermore, there is a few hundredths of a magnitude variation in the amplitude over the course of a number of periods in the 2D light curve, suggesting that some fundamental mode contamination remains. To make a better comparison, we recomputed the 2D calculation imposing the first overtone velocity distribution instead of the fundamental mode velocity distribution on the static model. The result is that the 2D amplitude is reduced from about 0.50 mag to 0.46 mag, but still noticeably higher than the 3D 0.38 mag. This suggests a genuine difference (see Figure~\ref{f4}). One can see that the two light curves have nearly the same shape; it is just the amplitude which is different.

To explore the nature of this difference in amplitude we have considered the following possibilities: it is a physical difference related to how convection alters pulsation in 2D and 3D, it is related to how large eddy simulations differ in 2D and 3D, and it is a numerical effect perhaps related to the different zonings in 2D and 3D. For the first of these we will examine the relationship between the light and velocity curves and the time dependent convective behaviour on pulsation phase. The second we investigate by examining the effects of eddy viscosity coefficient variation in 2D and 3D. Finally we study the effects of angular zoning in 2D calculations.

\begin{figure}
\center
\plotone{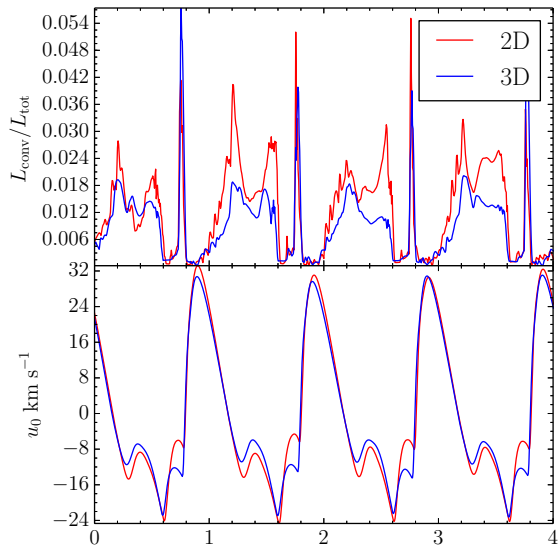}
\caption{Comparison of the phase dependence of the convective luminosity in 2D and 3D(top) for the first overtone 6900~K effective temperature model. The surface pulsational velocity is shown on the bottom.}
\label{f5}
\end{figure}

To explore if the differences in amplitude are physical in origin we compared the surface pulsational velocities of the two models and found the amplitudes to be quite close, to within about 1 km s$^{-1}$ out of a total amplitude of approximately 55 km s$^{-1}$.  This suggests a difference between the 2D and 3D models that alters the relationship between light and velocity amplitudes a modest amount. In both models the convective flux is always small, as seen in the time dependent history of the maximum ratio of the convective to total luminosity shown in Figure~\ref{f5}. We see that the time dependence is similar for both 2D and 3D, with the primary difference between the two being the larger ratio of the convective luminosity to the total luminosity in 2D just prior to the time of maximum contraction velocity (approximately phases 0.4 to 0.6 in Figure~\ref{f5}). At this time there appears to be a little more mass in the hydrogen and first helium ionization zones in 2D than in 3D, which may produce a larger amount of driving in 2D than 3D, although the discrete nature of the zoning and the defining of the boundaries of the ionization region makes this conclusion somewhat less than compelling. The other difference of note is that the velocity at the second local maximum near the end of contraction is closer to zero in 2D than in 3D (approximately phase 0.75) when the large spike in convective luminosity is seen in both 2D and 3D. It is not clear how this relates to the difference in light amplitude if at all.

We previously found that increasing the free constant in the eddy viscosity coefficient by a factor of two decreased the amplitude of the light curve by about 10\%.  We have performed similar calculations with both a 2D and 3D model and achieved a similar rate of change for both, matching our previous result, although there are some exceptions. In addition to the free constant the eddy viscosity coefficient depends on a characteristic length scale which is equated with the grid zoning (see equation~15 of GD2). In 2D we have taken this length scale to be the square root of the product of the zone sizes in the two directions. In 3D it is the cube root of the product of the zone sizes in the three directions. Because the radial zoning is generally finer than the angular zoning, the characteristic length scale tends to be slightly larger in 3D than in 2D. This operates in the direction of making the 3D light curve amplitude smaller than it would be if we used the 2D characteristic length scale. While this is not sufficiently large to explain the difference for the 6900~K model, it may be compounded by the fact that the larger amplitude variation will compress the radial zoning further, making the eddy viscosity even less. We summarize these results in Table~\ref{t2}. The table includes the value of the eddy viscosity parameter (0.17) used in all our other calculations. We see that velocity amplitude increases as we increase the eddy viscosity coefficient in both 2D and 3D (indeed, the velocity amplitude appears to be the same in 2D and 3D). However, the light amplitude decreases in the 3D case, while it is basically the same in the 2D case. This suggests that the light amplitude is a fairly sensitive feature in the calculations. Possibly a very broad parameter study, beyond the range of this current work, may be required to resolve this.

While these differences in the light curves are noticeable, it might be regarded as more surprising that they are not larger than they are, given that the 2D convective flow patterns are different from the 3D patterns. The primary difference is that in 2D the downward flow is in a trench rather than the relatively narrow column in 3D. This difference makes it somewhat difficult to compare 2D convection results such as the convective flux with the 3D analogue (see discussion in GD3). Another way to change the characteristic length scale is to change the grid zone size in the calculation. While extensive calculations in 3D are computationally prohibitive, we can perform some in 2D and present the results in the next section.

\section{ANGULAR RESOLUTION STUDY}
\label{sec:phase-dep-con-flux}

\begin{figure}
\center
\plotone{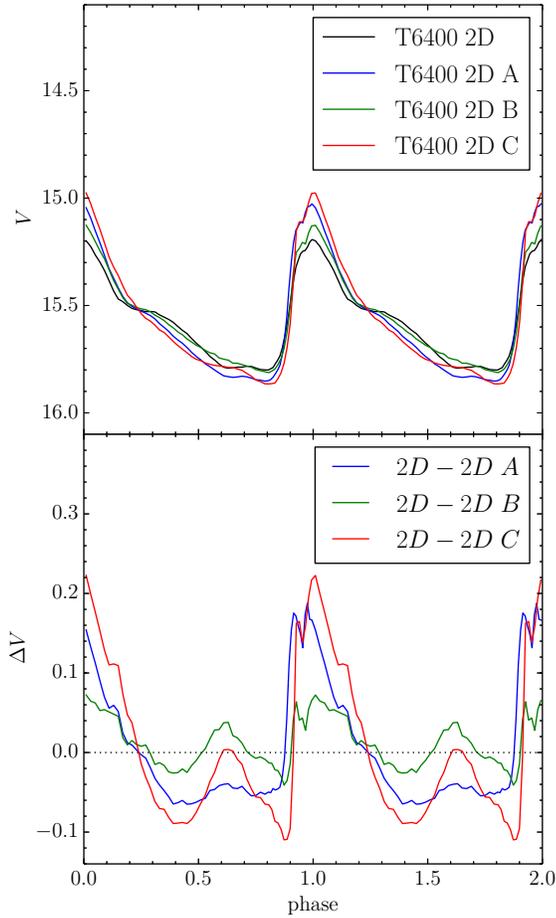}
\caption{Comparison of the light curves for Case A (same horizontal extent and twice the angular resolution),Case B (twice the horizontal extent and the same angular resolution), Case C (twice the horizontal extent and angular resolution) and the baseline calculation.}
\label{f6}
\end{figure}

\begin{figure}
\center
\plotone{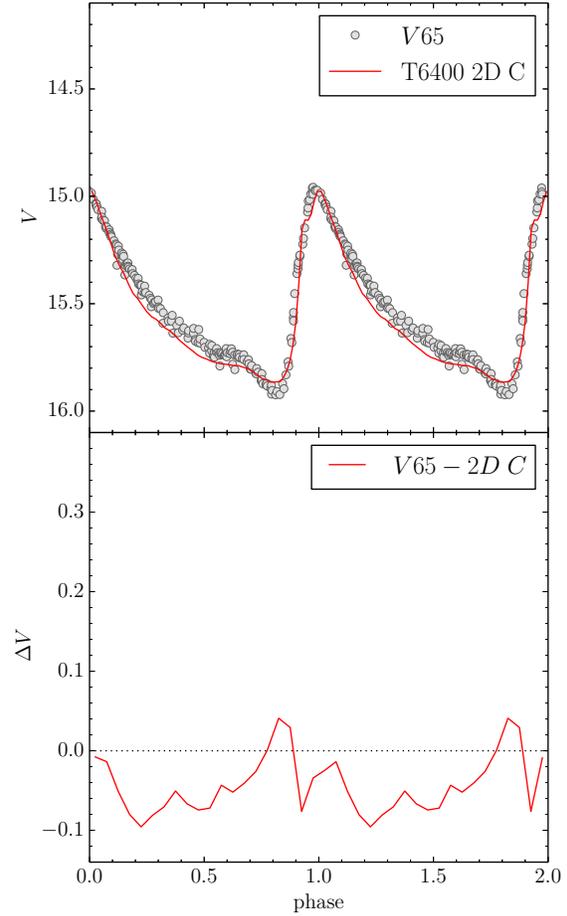}
\caption{Comparison of the light curves of Case C (twice the horizontal extent and angular resolution) and V65 in M3.}
\label{fig:Case-B}
\label{f7}
\end{figure}

\begin{deluxetable}{cccccccccccc}
\tabletypesize{\small}
\tablecaption{Angular Resolution Study of 6400~K Model}
\tablenum{3}
\tablehead{
\colhead{Case} &
\colhead{zones} &
\colhead{extent} &
\colhead{conv. cells} &
\colhead{$A_V$} &
\colhead{$L_{\rm conv.}/L_{\rm tot.}$} &
\colhead{$\phi_L$} &
\colhead{$\Delta \langle T\rangle / \langle T\rangle$} &
\colhead{$\phi_T$} &
\colhead{$v_{\rm conv.}$} &
\colhead{$\phi_v$} &
\colhead{$v_{\rm amp.}$} \\
\colhead{} &
\colhead{} &
\colhead{} &
\colhead{} &
\colhead{mag} &
\colhead{} &
\colhead{} &
\colhead{} &
\colhead{} &
\colhead{km s$^{-1}$} &
\colhead{} &
\colhead{km s$^{-1}$}
}
\startdata
Baseline &20& 6$^{\circ}$&1&0.64&$0.65\pm 0.01$&$0.73$&$0.65\pm 0.02$&$0.76$&$20\pm 2$&$0.71$&$82 \pm 1$\\
A        &40& 6$^{\circ}$&1&0.83&$0.61\pm 0.04$&$0.74$&$0.68\pm 0.01$&$0.75$&$25\pm 2$&$0.73$&$91 \pm 3$\\
B        &40&12$^{\circ}$&2&0.78&$0.64\pm 0.03$&$0.73$&$0.68\pm 0.01$&$0.76$&$31\pm 1$&$0.72$&$88 \pm 1$\\
C        &80&12$^{\circ}$&2&0.94&$0.56\pm 0.05$&$0.75$&$0.69\pm 0.01$&$0.76$&$31\pm 3$&$0.74$&$95 \pm 1$\\
\enddata
\label{t3}
\end{deluxetable}

We previously found that doubling the radial zoning did not appreciably alter either the growth rate or the full amplitude light curve (see GD2). However, until now we have not examined the angular zoning. The long time to completion makes this impractical for 3D calculations, so we focus on the latitudinal zoning for our 2D calculations. To this end we have performed a set of three new full amplitude calculations for $T_{\rm eff}=6400$~K with different angular zonings for comparison with the baseline calculations above: 1) calculation with the same horizontal extent but twice the angular resolution (case A), 2) calculation with twice the horizontal extent but the same angular resolution (case B), 3) calculation with twice the horizontal extent and twice the angular resolution (case C). The same properties as in Table~\ref{t1} are summarized in Table~\ref{t3} for the baseline case and these three angular zoning variations. We see that both light and velocity amplitudes generally increase as the resolution gets better and the extent gets larger. The horizontal temperature amplitude and phase do not appear to be much affected, while the convective velocities appear larger in the wider extent calculations.

Visual inspection of the convective flow patterns of the baseline calculation compared to cases A, B, and C show that the larger extent cases (B and C) contain two convective cells, while the smaller extant cases (baseline and A) contain only one convective cell. The appearance of the large scale flow patterns remain the same between the baseline case and case A and between the larger extent calculation cases (B and C), in that a convective cell is composed of one fast narrow down-flow and a slow broad up-flow (an example of such a flow is given in Figures~1 and 2 in GD3); there is little difference in the large scale structure of the convective flow patterns from lower resolution to higher resolution.

The comparison between light curves of the baseline calculation and cases A, B and C are presented in Figure~\ref{f6}. We see that the shape of the light curve is not very different although the amplitude is approximately 0.2 mag larger for the higher resolution case (Case A). Perhaps this should be expected because higher angular resolution makes the eddy length scale smaller, and we have already shown that reducing the eddy viscosity coefficient increases the amplitude (see discussion above).

The shape of the light curve does change in case B, particularly near minimum light. The prominent stand still in the baseline case at about phase 0.6 has been almost eliminated. Again the pulsation amplitude is increased with respect to the baseline case, but only by about 0.1 mag. 

The amplitude of the light curve in case C is larger than either case A or case B, perhaps suggesting an additive effect. The primary difference to the shape of the light curve is the loss of the long stand still at minimum light, seen in both case A and the baseline case, but with a remnant of a stand still prior to the final dip to minimum light which might be seen as a composite of case A and case B. The near stand still during falling light near mean light present in the baseline case and in case B and less prominently in case A is virtually absent in case C. However, we have noticed that a number of detailed features such as this in light curves are amplitude dependent so that ascribing it to a particular resolution must be treated with caution. We compare the light curves of case C and v65 in Figure~\ref{f7} were it is clear that they have features in common, such as the large dip at minimum light and the near stand still before that dip. However, we note that the model at a given phase during declining light is always less luminous. Despite the difference in amplitude, we find that the pulsational growth rates are essentially the same for all four cases.

We have computed full amplitude solutions for case C for effective temperatures from 6300-6900 K for comparison with the baseline case. As the effective temperature increases the change in amplitude between the baseline case and case C decreases from about 0.3 mag for $T_{\rm eff}=6400$~K to about 0.1 mag for $T_{\rm eff}=6800$~K. The change is only about 0.05 mag for the first overtone $T_{\rm eff}=6900$~K case (see Figure~\ref{f1}). This decrease in effect with increasing effective temperature is most likely due to decreasing importance of convection.

Based on all these results, it would appear that the amplitude and some aspects of the shape of the light curve are relatively sensitive to the various numerical aspects of the model. This should not disguise the fact that the time dependent interaction between convection and pulsation is clear in general with only relatively minor differences in detail.

\section{FINAL COMMENTS}
Over the past three years we developed a code capable of simulating the effects of convection on pulsation in RR Lyrae stars. The code can be run in one, two, or three dimensions.  A key to allowing us to compute full amplitude RR Lyrae models is forcing the radial zoning to move in such a way that the mass in any given spherical shell is constant throughout the evolution. This does not mean that the flow is Lagrangian, although it is in one dimension, but that there is no net flow of mass in or out of a spherical shell during the calculation.

The convective flow patterns are of course different in 2D and 3D, but the dependence of the convective flux on pulsation phase is quite similar. This suggests that 2D calculations are not a bad surrogate for 3D calculations, a point emphasized by the slight differences in pulsational growth rates.  The light curves are reasonable representations of those observed, and those near the red edge are quite a bit better than calculated with 1D models using a time dependent mixing length theory. 

This does not mean that all issues are resolved. There are differences between the light curves in the 2D and 3D calculations of the same RR Lyrae model, and the amplitude of the pulsation is rather sensitive to parameters of the eddy viscosity treatment, particularly the eddy viscosity coefficient and the characteristic length scale (here assumed to be related to the size of the computational mesh). At least part of this sensitivity probably arises from the fact that the zoning is still too coarse for the turbulent cascade to extend to sufficiently small scales. It is interesting that the 2D light curves generally look more like the observed light curves than the 3D light curves. We have presented some zoning studies in 2D that suggest that finer zoning is needed in the angular directions.  Because it takes several tens of millions of time steps to compute a full amplitude model, even with a sizeable initial pulsational velocity, the likelihood of being able to compute models with significantly better angular zoning in 3D is remote, although it would be feasible to double our angular zoning again (to 160 zones) for a limited number of models in 2D. Another generation or two of computing power will probably allow more finely resolved 3D calculations to be made, although they will still be time consuming for the foreseeable future.

\acknowledgments
Many of these calculations would not have been possible without the support of Compute Canada, and particularly of ACEnet, the high performance computing provider in Atlantic Canada. ACEnet is funded by the Canada Foundation for Innovation and provincial funding agencies of Nova Scotia, New Brunswick, and Newfoundland and Labrador. CMG received partial financial support during writing and analysis from a Consolidated STFC grant (ST/J001627/1). Persons potentially interested in becoming users of the SPHERLS code should contact CMG.

\bibliographystyle{apj}
%\bibliography{../../References/references}

\begin{thebibliography}{33}
\expandafter\ifx\csname natexlab\endcsname\relax\def\natexlab#1{#1}\fi

\bibitem[{{Alexander} \& {Ferguson}(1994)}]{Alexander-1994}
{Alexander}, D.~R., \& {Ferguson}, J.~W. 1994, ApJ, 437, 879

\bibitem[{{Bono} {et~al.}(1997{\natexlab{a}}){Bono}, {Caputo}, {Cassisi},
  {Incerpi}, \& {Marconi}}]{Bono-1997a}
{Bono}, G., {Caputo}, F., {Cassisi}, S., {Incerpi}, R., \& {Marconi}, M.
  1997{\natexlab{a}}, \apj, 483, 811

\bibitem[{{Bono} {et~al.}(1997{\natexlab{b}}){Bono}, {Caputo}, {Castellani}, \&
  {Marconi}}]{Bono-1997b}
{Bono}, G., {Caputo}, F., {Castellani}, V., \& {Marconi}, M.
  1997{\natexlab{b}}, \aaps, 121, 327

\bibitem[{{Bono} \& {Stellingwerf}(1994)}]{Bono-1994}
{Bono}, G., \& {Stellingwerf}, R.~F. 1994, \apjs, 93, 233

\bibitem[{{Buchler}(2009)}]{Buchler-2009}
{Buchler}, J.~R. 2009, in American Institute of Physics Conference Series, Vol.
  1170, American Institute of Physics Conference Series, ed. {J.~A.~Guzik \&
  P.~A.~Bradley}, 51--58

\bibitem[{{Cacciari} {et~al.}(2005){Cacciari}, {Corwin}, \&
  {Carney}}]{Cacciari-2005}
{Cacciari}, C., {Corwin}, T.~M., \& {Carney}, B.~W. 2005, \aj, 129, 267

\bibitem[{{Christy}(1966)}]{Christy-1966a}
{Christy}, R.~F. 1966, ARA\&A, 4, 353

\bibitem[{{Cox} {et~al.}(1966){Cox}, {Cox}, {Olsen}, {King}, \&
  {Eilers}}]{Cox-1966b}
{Cox}, J.~P., {Cox}, A.~N., {Olsen}, K.~H., {King}, D.~S., \& {Eilers}, D.~D.
  1966, \apj, 144, 1038

\bibitem[{{Deupree}(1977{\natexlab{a}})}]{Deupree-1977a}
{Deupree}, R.~G. 1977{\natexlab{a}}, ApJ, 211, 509

\bibitem[{{Deupree}(1977{\natexlab{b}})}]{Deupree-1977b}
---. 1977{\natexlab{b}}, \apj, 214, 502

\bibitem[{{Gastine} \& {Dintrans}(2008{\natexlab{a}})}]{Gastine-2008a}
{Gastine}, T., \& {Dintrans}, B. 2008{\natexlab{a}}, \aap, 484, 29

\bibitem[{{Gastine} \& {Dintrans}(2008{\natexlab{b}})}]{Gastine-2008b}
---. 2008{\natexlab{b}}, \aap, 490, 743

\bibitem[{{Gastine} \& {Dintrans}(2011)}]{Gastine-2011}
---. 2011, \aap, 528, A6

\bibitem[{{Gehmeyr}(1992{\natexlab{a}})}]{Gehmeyr-1992a}
{Gehmeyr}, M. 1992{\natexlab{a}}, \apj, 399, 265

\bibitem[{{Gehmeyr}(1992{\natexlab{b}})}]{Gehmeyr-1992b}
---. 1992{\natexlab{b}}, \apj, 399, 272

\bibitem[{{Gehmeyr}(1993)}]{Gehmeyr-1993}
---. 1993, \apj, 412, 341

\bibitem[{{Geroux} \& {Deupree}(2011)}]{Geroux-2011}
{Geroux}, C.~M., \& {Deupree}, R.~G. 2011, \apj, 731, 18

\bibitem[{{Geroux} \& {Deupree}(2013)}]{Geroux-2013b}
---. 2013, \apj, 771, 113

\bibitem[{{Geroux} \& {Deupree}(2014)}]{Geroux-2014}
---. 2014, \apj, 783, 107

\bibitem[{{Iglesias} \& {Rogers}(1996)}]{Iglesias-1996}
{Iglesias}, C.~A., \& {Rogers}, F.~J. 1996, ApJ, 464, 943

\bibitem[{{Kuhfuss}(1986)}]{Kuhfuss-1986}
{Kuhfuss}, R. 1986, A\&A, 160, 116

\bibitem[{{Marconi}(2009)}]{Marconi-2009}
{Marconi}, M. 2009, in American Institute of Physics Conference Series, Vol.
  1170, American Institute of Physics Conference Series, ed. {J.~A.~Guzik \&
  P.~A.~Bradley}, 223--234

\bibitem[{{Marconi} {et~al.}(2003){Marconi}, {Caputo}, {Di Criscienzo}, \&
  {Castellani}}]{Marconi-2003}
{Marconi}, M., {Caputo}, F., {Di Criscienzo}, M., \& {Castellani}, M. 2003,
  ApJ, 596, 299

\bibitem[{{Marconi} \& {Degl'Innocenti}(2007)}]{Marconi-2007}
{Marconi}, M., \& {Degl'Innocenti}, S. 2007, A\&A, 474, 557

\bibitem[{{Mundprecht} {et~al.}(2013){Mundprecht}, {Muthsam}, \&
  {Kupka}}]{Mundprecht-2013}
{Mundprecht}, E., {Muthsam}, H.~J., \& {Kupka}, F. 2013, \mnras, 435, 3191

\bibitem[{{Rogers} {et~al.}(1996){Rogers}, {Swenson}, \&
  {Iglesias}}]{Rogers-1996}
{Rogers}, F.~J., {Swenson}, F.~J., \& {Iglesias}, C.~A. 1996, ApJ, 456, 902

\bibitem[{{Stellingwerf}(1982{\natexlab{a}})}]{Stellingwerf-1982a}
{Stellingwerf}, R.~F. 1982{\natexlab{a}}, \apj, 262, 330

\bibitem[{{Stellingwerf}(1982{\natexlab{b}})}]{Stellingwerf-1982b}
---. 1982{\natexlab{b}}, \apj, 262, 339

\bibitem[{{Stellingwerf}(1984{\natexlab{a}})}]{Stellingwerf-1984a}
---. 1984{\natexlab{a}}, \apj, 277, 322

\bibitem[{{Stellingwerf}(1984{\natexlab{b}})}]{Stellingwerf-1984b}
---. 1984{\natexlab{b}}, \apj, 277, 327

\bibitem[{{Stellingwerf}(1984{\natexlab{c}})}]{Stellingwerf-1984c}
---. 1984{\natexlab{c}}, \apj, 284, 712

\bibitem[{{Tuggle} \& {Iben}(1973)}]{Tuggle-1973}
{Tuggle}, R.~S., \& {Iben}, I.~J. 1973, \apj, 186, 593

\bibitem[{{Xiong}(1989)}]{Xiong-1989}
{Xiong}, D. 1989, \aap, 209, 126

\end{thebibliography}

\end{document}